\documentclass[twocolumn,showpacs,prl,letter]{revtex4}

\usepackage{amsmath,amssymb,amsthm}

\usepackage{epsfig}

\usepackage{graphicx}

\begin{document}

\title{Alice falls into a black hole: Entanglement in non-inertial frames}
\author{I. Fuentes-Schuller\footnote{Published before under name Fuentes-Guridi}$^{\diamond}$$^\ddag$, R.
B. Mann$\dagger$$\ddag$} \affiliation{$^{\diamond}$ Centre for
Quantum Computation, Clarendon Laboratory,
University of Oxford, Parks Road OX1 3PU \\
$\dagger$ Department of Physics, University of Waterloo, Waterloo,
Ontario
Canada N2L 3G1\\
$^\ddag$Perimeter Institute, 31 Caroline Street North Waterloo,
Ontario Canada N2L 2Y5}

\begin{abstract}

Two observers determine the entanglement between two free bosonic
modes by each detecting one of the modes and observing the
correlations  between their measurements. We show that a state
which is maximally entangled in an inertial frame  becomes  less
entangled if the observers are relatively accelerated. This
phenomenon, which is a consequence of the Unruh effect, shows that
entanglement is an observer-dependent quantity in non-inertial
frames. In the high acceleration limit, our results can be applied
to a non-accelerated observer falling into a black hole while the
accelerated one barely escapes. If the observer escapes with
infinite acceleration, the state's distillable entanglement
vanishes.
\end{abstract}


\maketitle

Entanglement is a property of multipartite quantum states that
arises from the tensor product structure of the Hilbert space and
the superposition principle. It is considered to be a resource for
quantum information tasks such as teleportation
\cite{teleportation} and has applications in quantum control
\cite{control} and quantum simulations \cite{simulations}.
Non-relativistic bipartite entanglement can be quantified uniquely
for pure states by the von Neumann entropy and for mixed states
several measures have been proposed such as entanglement cost,
distillable entanglement and logarithmic negativity
\cite{entanglement}. Understanding entanglement in the
relativistic framework is crucial from both fundamental and
practical perspectives. Relativistic space-time presents naturally
a more complete setting for theoretical considerations and many
experimental set-ups require such a treatment. This program is
therefore an important and topical one. It is only in this
framework that we can understand  quantum information  tasks
involving entanglement between moving observers. A central
question in the field of relativistic quantum information is
whether entanglement is observer-independent. So far, it has been
shown that entanglement between inertial moving parties remains
constant although the entanglement between some degrees of freedom
can be transferred to others \cite{inertial}.

In this letter we investigate the entanglement between  two modes
of a non-interacting  massless scalar field when  one of  the
observers describing the state is uniformly accelerated. We
consider a maximally entangled pure state in an inertial frame and
describe its entanglement from a non-inertial perspective. Our
results imply that only inertial observers in flat spacetime agree
on the degree of entanglement, whereas non-inertial observers see
a degradation. While Minkowski coordinates $(t,z)$ are the most
suitable to describe the field from an inertial perspective,
Rindler coordinates $(\tau ,\xi )$ are  appropriate for discussing
the viewpoint of an observer  moving  with uniform acceleration.
Two different sets of Rindler coordinates, which differ from each
other by a sign change in the temporal coordinate, are necessary
for covering Minkowski space. These sets of coordinates define two
Rindler regions that are causally disconnected from each other. A
particle undergoing uniform acceleration in a given Rindler region
remains constrained to it and has no access to the other Rindler
sector. The solutions of the Klein-Gordon equation for a massless
scalar field in Minkowski coordinates are related to the solutions
of the equation in Rindler coordinates through Bogoliubov
transformations. Using these transformations one finds that the
ground state of a given mode seen by an inertial observer in
Minkowski coordinates corresponds to a two-mode squeezed state in
Rindler coordinates \cite{walls}. These two modes respectively
correspond to  the field observed in the two distinct Rindler
regions. An observer moving with uniform acceleration in one of
the regions has no access to field modes in the causally
disconnected region. Therefore, the observer must trace over the
inaccessible region losing information about the state, which
essentially results in the detection of a thermal state. This is
known as the Unruh effect \cite{unruh}.

A consequence of this effect is that an entangled pure state seen
by inertial observers appears mixed from an accelerated frame. In
this case entropy no longer quantifies entanglement. However, it
is possible to
determine the entanglement of such a state using the logarithmic negativity %
which is a full entanglement monotone that bounds distillable
entanglement from above\cite{vidal} . In our analysis we use the
mutual information \cite{mutual} to quantify the state's total
correlations (classical plus quantum). It is interesting to note
that the Schwarzschild space-time very close to the horizon
resembles Rindler space in the infinite acceleration limit
\cite{black}. Therefore our technique can be applied to study the
entanglement between two scalar modes seen by observers near an
event horizon. We will see that when two modes of the field are
maximally entangled in an inertial frame, the presence of the
horizon degrades the entanglement seen by one observer falling and
the other escaping the fall into a black hole. The state remains
only classically correlated when the acceleration approaches
infinity. We prove this by showing that, in the infinite
acceleration limit in Rindler space, the logarithmic negativity is
zero.

To formalize the above, consider that two modes, $k$ and $s$, of a
free massless scalar field  in Minkowski spacetime are maximally
entangled from an inertial perspective, i.e., the quantum field is
in a state
\begin{equation}
\frac{1}{\sqrt{2}}\left( \left|
0_{s}\right\rangle^{\mathcal{M}}\left|0_{k}\right\rangle
^{\mathcal{M}}+\left| 1_{s}\right\rangle ^{\mathcal{M}}\left|
1_{k}\right\rangle ^{\mathcal{M}}\right) . \label{eq:minko}
\end{equation}%
The states $\left| 0_{j}\right\rangle ^{\mathcal{M}}$ and
$\left|1_{j}\right\rangle ^{\mathcal{M}}$ are the vacuum and
single particle excitation states of the mode $j$ in Minkowski
space. We assume that Alice has a detector which only detects mode
$s$ and Rob has a detector sensitive only to mode $k$. If Rob
undergoes uniform acceleration $a$, the states corresponding to
mode $k$ must be specified in Rindler coordinates in order to
describe what Rob sees. Considering only one spatial dimension
$z$, the world lines of uniformly accelerated observers in
Minkowski coordinates correspond to hyperbolae, to the left
(region I) and right (region II) of the origin, bounded by
light-like asymptotes constituting the Rindler horizon. The
Rindler coordinates are defined by
\begin{eqnarray}
t &=&a^{-1}e^{a\xi }\sinh {a\tau },\quad z=a^{-1}e^{a\xi }\cosh {a\tau },\quad |z|<t, \\
t &=&-a^{-1}e^{a\xi }\sinh {a\tau },\quad z=a^{-1}e^{a\xi }\cosh {a\tau }%
,\quad |z|>t,  \notag
\end{eqnarray}%
where the hyperbolae correspond to the space-like coordinates $\xi $ and $%
\tau $ is the proper time, i.e., the length of the hyperbolic
world line measured by the Minkowski metric. The Minkowski vacuum
state, defined as the absence of any particle excitation in any of
the modes
\begin{equation}
|0\rangle ^{\mathcal{M}}=\prod_{j}|0_{j}\rangle ^{\mathcal{M}},
\end{equation}%
can be expressed in terms of a product of two-mode squeezed states
of the Rindler vacuum, \cite{walls}
\begin{eqnarray}
\left| 0_{k}\right\rangle ^{\mathcal{M}} &\sim &\frac{1}{\cosh r}%
\sum_{n=0}^{\infty }\tanh ^{n}r\,\left| n_{k}\right\rangle
_{I}\left|n_{k}\right\rangle _{II},  \label{eq:vacuum} \\
\cosh r &=&(1-e^{-2\pi \Omega })^{-1/2},\quad \Omega =|k|c/a.
\end{eqnarray}%
where  $\left| n_{k}\right\rangle _{I}$ and $\left|
n_{k}\right\rangle _{II}$ refer to the mode decomposition in
region I and II, respectively, of Rindler
space. Each Minkowski mode $j$ has a Rindler mode expansion given by Eq. (%
\ref{eq:vacuum}). In our problem, we consider detectors sensitive
to a single Minkowski mode $s$ for Alice and $k$ for Rob and we
consider that the rest of the modes in the field are in the
vacuum. In our analysis we trace over all the modes except for $s$
and $k$. The result of this trace is a pure state because
different modes $j$ and $j^{\prime }$ do not mix.

Using Eq. (\ref{eq:vacuum}) and
\begin{equation}
\left| 1_{k}\right\rangle ^{\mathcal{M}}=\frac{1}{\cosh ^{2}r}%
\sum_{n=0}^{\infty }\tanh ^{n}r\,\sqrt{n+1}\left|
(n+1)_{k}\right\rangle _{I}\left| n_{k}\right\rangle _{II},
\notag
\end{equation}%
we can rewrite Eq.(\ref{eq:minko}) in terms of Minkowski modes for
Alice and Rindler modes for Rob. Since Rob is causally
disconnected from region II, we must trace over the states in this
region, which results in a mixed state
\begin{eqnarray}
\rho _{AR} &=&\frac{1}{2\cosh ^{2}r}\sum_{n}(\tanh r)^{2n}\rho _{n}, \\
\rho _{n} &=&|0n\rangle \langle 0n|+\frac{\sqrt{n+1}}{\cosh
r}|0n\rangle\langle 1n+1|  \notag \\
&+&\frac{\sqrt{n+1}}{\cosh r}|1n+1\rangle \langle
0n|+\frac{\left(n+1\right) }{\cosh ^{2}r}|1n+1\rangle \langle
1n+1|  \notag
\end{eqnarray}%
where $|nm\rangle =|n_{s}\rangle ^{\mathcal{M}}|m_{k}\rangle
_{I}$. The partial transpose criterion \cite{peres} provides a
sufficient criterion for entanglement. If at least one eigenvalue
of the partial transpose is negative, then the density matrix is
entangled; but a state with positive partial transpose can still
be entangled. This type of entanglement is called bound or
non-distillable entanglement \cite{vidal}. We obtain the partial
transpose by interchanging Alice's qubits and we find the
eigenvalues in the $\left(n,n+1\right) $ block to be
\begin{equation*}
\lambda _{\pm }^{n}=\frac{\tanh ^{2n}r}{\left( 4\cosh ^{2}r\right)
}\left[ \left( \frac{n}{\sinh ^{2}r}+\tanh ^{2}r\right) \pm
\sqrt{Z_{n}}\right] ,
\end{equation*}
where
\begin{equation}
Z_{n}=\left( \frac{n}{\sinh ^{2}r}+\tanh ^{2}r\right)
^{2}+\frac{4}{\cosh ^{2}r}.  \notag
\end{equation}

\begin{center}
\begin{figure}[tbp]
\epsfig{file=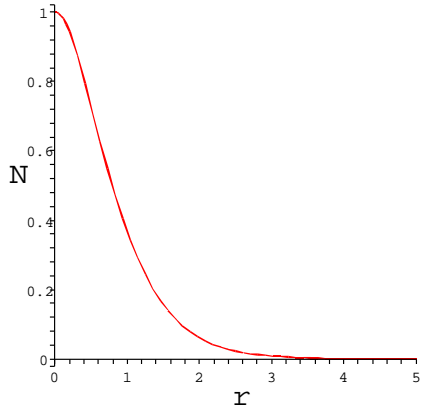,width=6cm} \caption{The negativity as
a function of the acceleration $r$.} \label{fig:neg}
\end{figure}
\end{center}

\begin{center}
\begin{figure}[tbp]
\epsfig{file=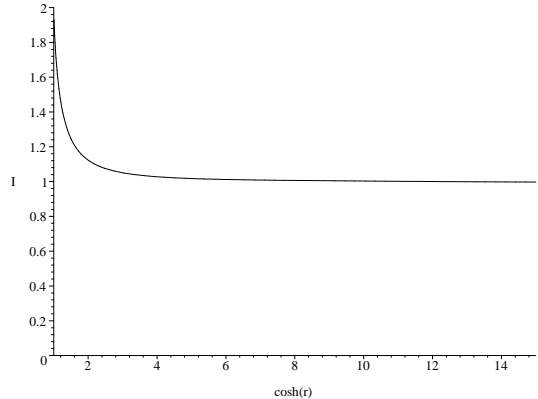, angle=-90, width=8cm} \caption{Mutual
information as a function of $\cosh (r)$.} \label{fig:mutual}
\end{figure}
\end{center}

It is clear that for finite acceleration $(r<\infty )$ one
eigenvalue is always negative; thus the state is always entangled.
Only in the limit $r\rightarrow \infty $ could the negative
eigenvalue possibly go to zero. To investigate this further, we
sum over all the negative eigenvalues and calculate the
logarithmic negativity. This entanglement monotone is defined as
$N(\rho )=\log _{2}||\rho ^{T}||_{1}$ where $||\rho ^{T}||_{1}$ is
the trace-norm of the density matrix $\rho $. The result is
$N(\rho _{AR})=\log _{2}(\frac{1}{2\cosh ^{2}{r}}+\Sigma )$ where
\begin{equation}
\Sigma =\sum_{n=0}^{\infty }\frac{\tanh ^{2n}r}{2\cosh
^{2}r}\sqrt{\left( \frac{n}{\sinh ^{2}r}+\tanh ^{2}r\right)
^{2}+\frac{4}{\cosh ^{2}r}}.  \notag \label{eq:sum}
\end{equation}%
For vanishing acceleration $(r=0)$, $N(\rho _{AR})=1$ as expected.
For finite acceleration the entanglement is degraded (fig
\ref{fig:neg}). The limit $r\rightarrow \infty $ can be explored
by analyzing an upper and lower bound on the negativity
constructed by bounding the sum in the above equation by two sums
that can be carried out exactly. We find
\begin{equation*}
1\leq \Sigma <\frac{2\cosh ^{2}r+2\cosh r}{2\cosh ^{2}r}.
\end{equation*}%
Since the bounds converge to $1$, the negativity is exactly $0$ in the limit.%
This means that the state has no longer distillable entanglement.
We can also estimate the total amount of correlation in the state
by calculating the mutual information, defined as
$I(\rho_{AR})=S(\rho _{A})+S(\rho _{R})-S(\rho _{AR})$ where
$S(\rho )=-\text{Tr}(\rho \log _{2}(\rho ))$ is the entropy of the
density matrix $\rho $. The entropy of the joint state is
\begin{eqnarray}
S(\rho _{AR}) &=&-\frac{1}{2\cosh ^{2}r}\sum_{n=0}^{\infty }\tanh ^{2n}r%
\text{L}_{n}, \\
\text{L}_{n} &=&\left( 1+\frac{n+1}{\cosh ^{2}r}\right) \log
_{2}\left[\frac{{\tanh ^{2n}r}}{2\cosh ^{2}r}\left( 1+\frac{n+1}{\cosh ^{2}r}\right) %
\right] .  \notag
\end{eqnarray}%
We obtain Rob's density matrix in region I by tracing over Alice's
states; its entropy is
\begin{eqnarray}
S(\rho _{RI}) &=&-\frac{1}{2\cosh ^{2}r}\sum_{n=0}^{\infty }\tanh ^{2n}r%
\text{M}_{n} \\
\text{M}_{n} &=&(1+\frac{n}{\sinh ^{2}r})\log _{2}{\frac{\tanh ^{2n}r}{%
2\cosh ^{2}r}(1+\frac{n}{\sinh ^{2}r})}.  \notag
\end{eqnarray}
Tracing over Rob's states we find Alice's density matrix:
\begin{equation}
\rho _{A}^{\mathcal{M}}=\frac{1}{2}(|0\rangle \langle 0|+|1\rangle
\langle 1|),
\end{equation}%
whose entropy is $S(\rho _{A})=1$. The mutual information is
\begin{eqnarray*}
I\left( N\right) &=&1-\frac{1}{2}\log _{2}\left( \tanh ^{2}r\right) -\frac{1}{2\cosh ^{2}r}\sum_{n=0}^{N}\tanh ^{2n}r\mathcal{D}_{n}, \\
\mathcal{D}_{n} &=&(1+\frac{n}{\sinh ^{2}r})\log _{2}\left( {1+\frac{n}{\sinh ^{2}r}}\right) \\
&-&(1+\frac{n+1}{\cosh ^{2}r})\log _{2}\left( {1+\frac{n+1}{\cosh
^{2}r}}\right),
\end{eqnarray*}%
which we plot in Fig. (\ref{fig:mutual}). For vanishing
acceleration, the mutual information is $2$. As the acceleration
increases, it becomes smaller, converging to unity in the limit of
infinite acceleration. Note that a maximally mixed state of
maximally entangled states has mutual information equal to one.
Since the distillable entanglement in the infinite acceleration
limit is zero, we know that in this limit the total correlations
consist of classical correlations plus bound entanglement.
The entropy of the density matrices for Rob and Alice in region I
and Rob in region II are equal $S(\rho _{ARI})=S(\rho _{RII})$.
This is because the state in Eq. (\ref{eq:minko}) is pure, and
therefore the entropies of the reduced density matrices of any
bipartite division of the system are equal. In the limit of
infinite acceleration $S(\rho _{ARI})=1$. The modes in region II
are maximally entangled with the state in region I. When the
bosons are maximally entangled, for vanishing acceleration, there
is no distillable entanglement with region II. For finite
acceleration, the entanglement between the bosons is degraded as
the entanglement with region II grows. In general, entanglement in
tripartite pure states cannot be arbitrarily distributed amongst
the subsystems \cite{wooters}. This phenomenon, called
entanglement sharing, explains here why the entanglement between
the bosons is degraded as acceleration grows.

Our results for the infinite acceleration limit describe the
entanglement of the two bosonic modes seen by Alice and Rob in the
case that they are extremely close to the horizon of a  static
black hole. The Schwarzschild space-time describes the geometry of
space-time for a spherical non-rotating mass $m$. Considering only
the radial component, the metric is
\begin{equation}
ds^{2}=-\left( 1-\frac{2m}{R}\right) dT^{2}+\left(
\frac{1}{1-2m/R}\right) dR^{2}.
\end{equation}%
The presence of a Schwarzschild black hole corresponds to a region
causally cut off from the rest of spacetime by an horizon at
$R=2m$ . Changing coordinates so that $R-2m=x^{2}/8m$, we have
$1-2m/R=(Ax^{2})/(1+(Ax)^{2})\approx (Ax)^{2}$ near $x=0$ with
$A=1/4m$. This means that $dR^{2}=(Ax)^{2}$ and thus, very close
to the horizon of the black hole at $R\approx 2m$, the
Schwarzschild space-time can be approximated by Rindler space
\begin{equation}
ds^{2}=-(Ax)^{2}dT^{2}+dx^{2},
\end{equation}%
where the acceleration parameter $a=A^{-1}$. The infinite
acceleration limit corresponds to Rob moving on a trajectory
arbitrarily close to the Rindler horizon; in the context of a
black hole, this is arbitrarily close to the event horizon.
Therefore, our analysis can be applied to the case of Alice
falling into the black hole while Rob escapes. Each of them
measures one of the modes and Rob sends the results of his
experiment to Alice. Alice can then compare the results and
estimate the entanglement between the modes.

If we considered Alice to be accelerated as well, the density
matrix would be mixed to a higher degree, resulting in a higher
degradation of entanglement. Only two inertial observers  in that
space  would agree that the state investigated is maximally
entangled. This shows that entanglement is an  observer- dependent
quantity in non-inertial frames. The presence of a horizon for the
uniformly accelerated observers results in a loss of information
producing the degradation in the entanglement. In flat space-time
one could prescribe a well-defined notion of entanglement by
stating that only inertial observers are good observers of
entanglement. This is not a problem in this case since inertial
observers have a preferred role in flat spacetime. In curved
spacetime, even two nearby inertial observers are relatively
accelerated, due to the geodesic deviation equation. The results
of this paper strongly suggest that in curved spacetime not even
two inertial observers agree on the degree of entanglement of a
given bipartite quantum state of some quantum field. The detailed
analysis of  entanglement between modes of a quantum field on a
curved spacetime  however, is more involved, and will be treated
elsewhere \cite{ball}.

With the intention of investigating entanglement between
accelerated observers, the state fidelity in a teleportation
protocol was studied \cite{alsing} using relatively accelerated
cavities. It was found that the fidelity decreases as the
acceleration grows. Since state fidelity in conventional
teleportation protocols is related to entanglement, the authors
interpret this result as an indication of entanglement
degradation. Unfortunately, the mode expansions used in that work
corresponds to those of free space. Although  there is  some
indication that these  results are qualitativly correct, a
detailed calculation of the effects of an accelerated cavity still
remains to be done.

We have calculated the entanglement between two free modes of a
scalar field as seen by an inertial observer detecting one of the
modes and a uniformly accelerated observer detecting the second
mode. The entanglement which appeared to be maximal in an inertial
frame is then degraded by the Unruh effect. In the limit of
infinite acceleration, which can be applied to the situation of
one of the observers falling into a black hole while the other
barely escapes, the distillable entanglement vanishes but the
state remains correlated through classical correlations and bound
entanglement. The entanglement degradation between the bosons is
due to the increase of entanglement with the modes in the causally
disconnected Rindler region. The accelerated observer has only
partial access to the information and therefore entanglement
appears degraded. Similar effects have been noted to have
relevance for black hole entropy bounds \cite{Marolf}. A
well-defined notion of entanglement in flat space-time can be
provided by  restricting attention to inertial observers. In
curved spacetime, however, the notion of entanglement can be
expected to become a rather subtle one, as does the notion of
particles.

We especially thank Frederic P. Schuller for helpful discussions.
We gratefully acknowledge comments by Don Marolf, Martin Plenio,
Jonathan Ball and Daniel R. Terno. This work was supported in part
by the Natural Sciences and Engineering Research Council of
Canada.

%%%%%%%%%%%%%%%%%%%%%%%%%%%%%%%%%%%%%%%%%%%%%%%%%%%%%%%%%%%%%%%%%%%%%%%%%%%%%%%%%%%%%%%%%%%%%%%%%%%%

\end{document}